# Heroin addiction hijacks the Nucleus Accumbens: craving and reactivity to naturalistic stimuli

Greg Kronberg*, Ahmet O. Ceceli*, Yuefeng Huang, Pierre-Olivier Gaudreault, Sarah King, Natalie McClain, Pazia Miller, Lily Gabay, Devarshi Vasa, Pias Malaker, Defne Ekin, Nelly Alia-Klein, Rita Z. Goldstein
Icahn School of Medicine at Mount Sinai, New York, NY 10029
*equal contribution

**Abstract**

Drug-related cues hijack attention away from alternative reinforcers in drug addiction, inducing craving and motivating drug-seeking. However, the neural correlates underlying this biased processing, its expression in the real-world, and its relationship to cue-induced craving are not fully established, especially in opioid addiction. Here we tracked inter-brain synchronization in the Nucleus Accumbens (NAc), a hub of motivational salience, while heroin-addicted individuals and healthy control subjects watched the same engaging heroin-related movie. Strikingly, the left NAc was synchronized during drug scenes in the addicted individuals and non-drug scenes in controls, predicting scene- and movie-induced heroin craving in the former. Our results open a window into the neurobiology underlying shared drug-biased processing of naturalistic stimuli and cue-induced craving in opiate addiction as they unfold in the real world.

**Summary**

During an engaging movie, the NAc tracks drug scenes and predicts cue-induced craving in heroin addiction

**Main text**

Opioid-related overdose deaths in the United States have increased by 109% between 2015 and 2020, including a 38% increase in 2020 alone (*1*). Heroin use in particular increased by approximately 40% in response to the COVID-19 pandemic (*2*). Despite the magnitude of this drug use and overdose epidemic, however, and compared to other drugs of abuse, the neurobiological substrates of human opioid use disorder (OUD) have been largely understudied. In addiction to other drug classes (e.g., stimulants), drug cues have been shown to divert attention away from alternative reinforcers (e.g., social, food, sex) (*3*), a core imbalance in motivational salience suggested to drive drug use and perpetuate the addiction cycle (*4*). Whether a similar hijacking occurs in opioid addiction, however, remains an open question. Here we were interested in the extent (i.e., its expression as a phenotype shared by groups of individuals) of such drug-biased neural processing in real-world contexts and its relationship to cue-induced craving in heroin addiction.

To explore the brain substrates of cue reactivity and craving in human drug addiction, neuroimaging studies commonly employ static images (e.g., pictures of drugs). Instead, here we used an engaging heroin-related movie, a dynamic, narrative-based, and context-rich natural stimulus, to better approximate actual real-world experiences in individuals with OUD (iOUD) (*5*). Specifically, we used inter-subject synchronization (*6*) to analyze fMRI BOLD responses to the movie, hypothesizing a shared brain signature in iOUD. Our primary region of interest was the Nucleus Accumbens (NAc), a major processing hub for motivational salience attribution, reward anticipation, and craving (*7*, *8*), where we expected to document unique synchronized tracking of drug-related content as predictive of cue-induced craving in the iOUD.

Functional MRI BOLD activity was recorded in 29 treatment-seeking medically stabilized iOUD (40.4±10.3 years, 23 Male, 19 White) and 16 age, sex, and race-matched healthy controls (HC) (43.8±10.3 years, 10 M, 11 W), while subjects watched the first 17 minutes of the movie "Trainspotting", which contains scenes of explicit heroin use, as well as food and social scenes that are highly salient but not directly related to heroin use. Importantly, the overarching narrative structure of the movie is centered around drug-addiction, including complex contextual elements relevant to OUD (e.g., social, emotional, and economic challenges of addiction). To infer the movie features that led to synchronized NAc activity, we used an approach developed by Hasson et al. (*6*), which was inspired by reverse correlation analyses of single unit recordings (*9*, *10*). We were interested in whether the iOUD would show differential synchrony in the NAc during the drug-related content at the expense of other typically motivating stimuli.

Following Hasson et al., we decomposed the BOLD signal in each voxel during movie watching into two components. The first component, which we call the global component, relies on the observation that engaging narratives evoke a wave of global synchronized activity across much of the brain, driven by surprising and emotionally engaging moments of a movie (*6*). Here, the global component was derived by averaging the BOLD signal across all grey matter voxels and then z-scoring the resulting time series within each subject. The second component, which we call a selective component, was derived by regressing the global component out of the time series at each voxel, yielding a residual signal that is not explained by the global activation wave throughout the brain and is specific to each voxel. Hasson et al. has shown that a reverse correlation analysis of this selective component can recover canonical response properties from traditional GLM-based fMRI tasks [e.g., fusiform face area responds to scenes with faces (*6*)]. Here, for the first time, we applied this method for the purpose of tracking the neural signature of drug related motivation/incentive salience in subcortical regions in heroin-addicted individuals.

Applying this reverse correlation method to the mean selective component separately in the left and right NAc (LNAc and RNAc, respectively), we found a set of significant time points during the movie that elicited a synchronized NAc response in each subject group (see *Identifying synchronized TR's and region-specific reactivity* in supplemental methods). Accounting for the hemodynamic lag, we then identified clips of the movie that contributed to this synchronization (5-10 sec before each significant time point). Stitching these clips together yielded separate NAc-specific "movies" for each group. Remarkably, the LNAc-specific movie contained almost exclusively clips of drug use in the OUD group and no scenes of drug use in the HC group (Fig. 1; see supplemental videos).

For statistical confirmation purposes, and based on previous methods (*11*), we quantified the content of the LNAc-specific movies by first breaking the entire movie into 24 distinct scenes. Each scene was a priori classified as drug or non-drug, based on whether drugs or drug use were explicit in the scene (see supplemental methods section *Labeling of region-specific movies* for detailed criteria). Each TR from the LNAc-specific movie could then be matched to a scene (multiple TRs could be matched to the same scene) and thereby labeled as drug or non-drug, yielding a count of scene types to be used for statistical comparisons. Using this count as the dependent measure, chi-square tests revealed a significantly different scene type distribution between the iOUD and HC for the LNAc-specific movie, such that for iOUD it fell almost exclusively within the drug scenes, while for HC it was close to what could be expected if scenes were selected at random (iOUD: 40/8, HC: 25/40 drug/non-drug; $\chi^2(1)=20.95$, p=0.0000047). Indeed, nonparametric tests also indicated that the number of drug scenes was significantly different ($\alpha=0.05$) from random phase shuffled signals for iOUD, but not HC (Fig. S1). We further compared the magnitude of the LNAc-selective component directly between the two groups and found that it was significantly different during each of the LNAc-specific movies (Fig. S2).

Given the skewed responsiveness of the LNAc to drug stimuli in iOUD and this region's role in craving (*7*), we inspected whether the magnitude of the LNAc-selective component predicts cue-induced craving for drugs. In-scanner subjective heroin craving ratings were collected immediately before and after subjects watched the movie (for calculation of movie-induced craving). After leaving the scanner (within 45 minutes after watching the movie), subjects also completed a custom questionnaire intended to probe scene-specific cravings. The LNAc-specific component averaged over significant time points for each subject predicted both scene- and movie-induced cravings (Fig. 2A,B). Interestingly, the correlation with baseline craving (measured with the heroin craving questionnaire, HCQ) (*12*) was not significant (Fig. 2C), suggesting that the LNAc synchronous reactivity scales with the dynamic experience of craving induced by watching the movie.

We next tested whether the LNAc reactivity was related to other potential explanatory variables including the demographic measures that differed between the groups and, in iOUD, measures of recent and lifetime drug use, withdrawal symptoms and severity of dependence (Table S2). The LNAc reactivity signal was not significantly correlated ($\alpha=0.05$) with any of these measures, further suggesting that it was not driven by the demographic factors that commonly differ between individuals with and without addiction, and that it may specifically be predictive of craving induced by movie-watching. To better examine the specificity of the above results to the LNAc, we repeated the above analyses (including statistics and correlations) for the global component, several other regions of interest involved in drug cue reactivity [ventromedial prefrontal cortex (*13*), dorsolateral prefrontal cortex (*14*), anterior cingulate cortex (*15*), insula (*16*), putamen (*17*)], and a control region [fusiform cortex (*18*)]. Other than the LNAc, only the left anterior insula showed significant drug-biased reactivity in OUD (Table S3). However, only the LNAc showed a complete reversal of scene-type preferences between the groups (i.e., drug scenes for iOUD and non-drug scenes for HC) and a correlation with the cue-induced craving measures, suggesting specificity of these results to the LNAc.

Supporting the impaired response inhibition and salience attribution model, and other prominent neurobiological theories of addiction (*4, 19*), which posit an aberrant incentive salience attribution to drug-related stimuli at the expense of other typically motivating stimuli, we revealed a bias toward naturalistic depictions of drug-use in the LNAc of iOUD. Documenting this imbalance during exposure to a movie, a narrative-based and context-rich dynamic stimulus, suggests that such neural bias may dominate during real-life daily ongoing experience. Here we documented such a shared narrowing of responses in addiction towards drug-related stimuli even in currently abstinent, treatment-seeking and medication-maintained individuals. Predicting cue-induced craving, this interpretable brain signal may also be a powerful predictor of later drug use and relapse, as remains to be ascertained with longitudinal follow-up studies in these subjects.

Whether the dynamic bias of the NAc to a drug-related narrative is generalizable to other substance use disorders, or to individuals at risk for developing drug addiction, remains an open question. Future efforts may also benefit from more sex-balanced samples to interrogate potential sex-differences related to mapping drug cue reactivity and craving in a naturalistic context. Such efforts may also warrant larger samples, although the inter-subject correlation based statistics yield stable results with comparable samples in traditional block designs (*20*), and even smaller samples in naturalistic designs (*21*). Furthermore, future efforts could employ a whole-brain search for inter-brain synchronization during movie watching, complementing our regional focus (that accounted for the global component and corrected for multiple comparisons). A real-time measure of engagement during movie watching (e.g., with eye-tracking or online scene-specific ratings) may have provided more information about participants' experiences during the movie, but we decided in this first effort not to disrupt the continuous narrative structure and engagement during movie viewing, instead obtaining a comprehensive post-movie survey.

To the best of our knowledge, this is the first use of dynamic naturalistic stimuli for the demonstration of drug-biased processing as a predictor of cue-induced craving in the addicted human brain. The interbrain synchronization and scene classification methods introduced here for the first time in the context of heroin addiction allow us to highlight specific scenes that drive the shared striatal reactivity; as compared to typical block- or event-related picture-based studies, these methods can better decode the actual features/identity of the craving-inducing stimuli even in this treatment-seeking group. Crucially, this approach effectively operationalizes daily-life drug-related experiences, opening a window into the neurobiology underlying drug-biased processing and cue-induced craving as they unfold in the real world. As one iOUD referred to the movie stimulus "it gives me that feeling like even though I'm not getting high, I just, I feel like I'm doing it with them."

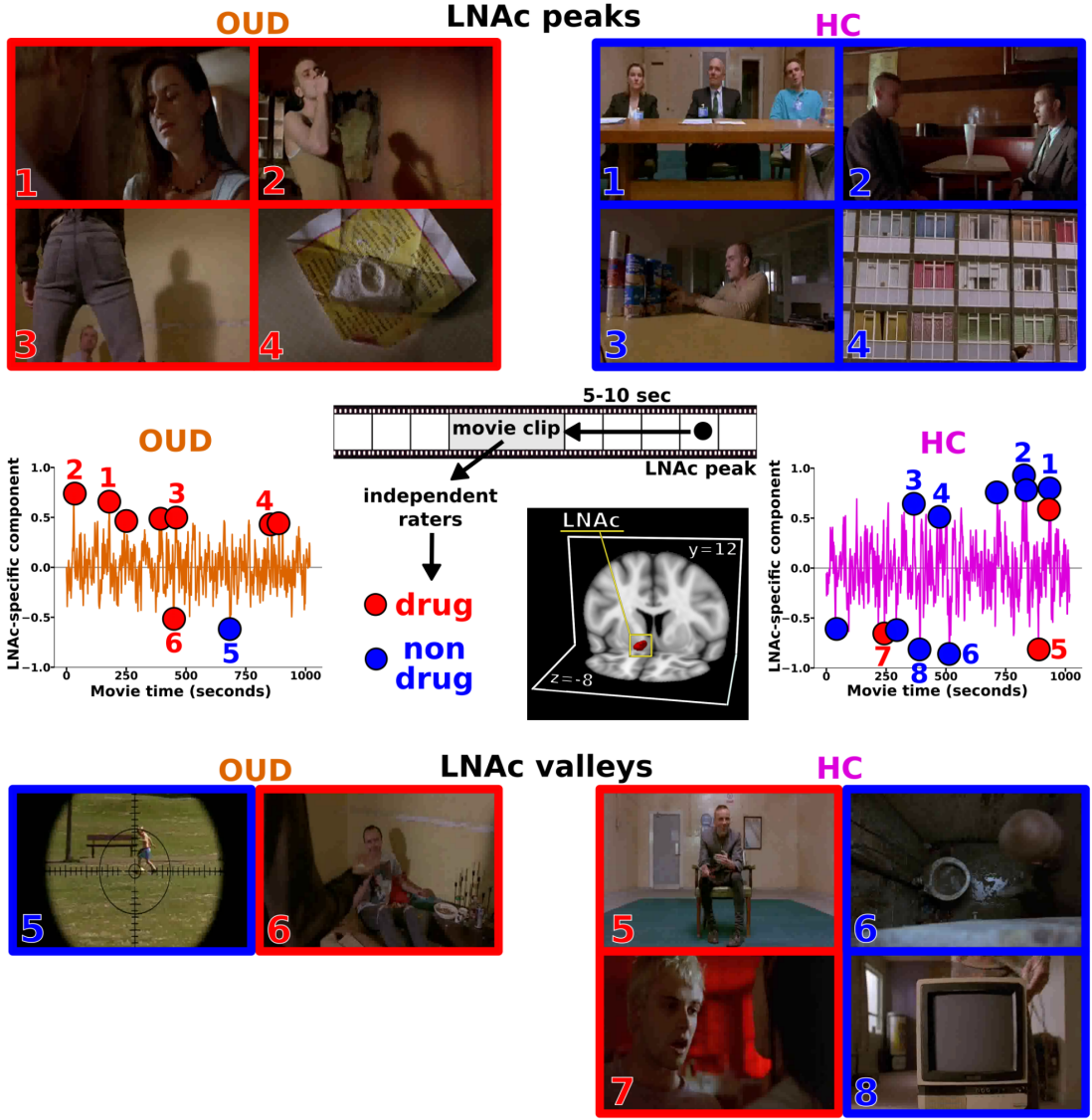

*Fig. 1. LNAc responds almost exclusively to drug use in OUD and non-drug scenes in HC*
The mean LNAc-specific component for each group is displayed with filled circles indicating significantly synchronized time points for each group (left: OUD; right: HC). Movie clips 5-10 s prior to each significant time point were labeled by independent raters as drug (red) or non-drug (blue). Representative still images are displayed for the four most significant peaks (top) and valleys (bottom) of the LNAc-specific component. While OUD responded almost exclusively to drug use, HC peaks corresponded to other typically motivating stimuli (e.g., food or social cues) and HC valleys to typically aversive stimuli (e.g., pain, embarrassment, feces). A 3 dimensional rendering of the LNAc mask (Harvard-Oxford Atlas) is displayed in MNI space.

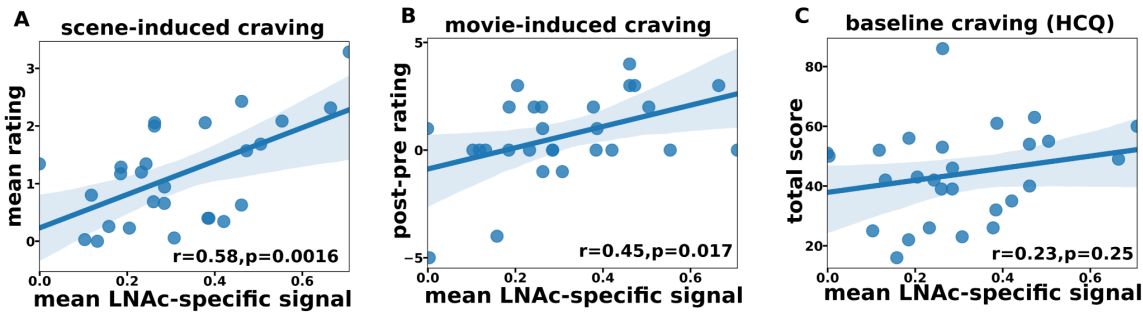

*Fig. 2. LNAc reactivity correlates with cue-induced heroin craving in OUD*
A. LNAc reactivity correlated with mean scene-induced craving ratings. These craving ratings were collected outside the scanner in response to 3 sec video clips from the movie (averaged over all clips). B. LNAc reactivity correlated with immediately post-minus-pre movie subjective ratings for heroin craving. C. LNac reactivity did not correlate with baseline craving as measured by the heroin craving questionnaire (HCQ). For all plots, outliers with a ±3 SD from the mean threshold criteria were removed. Line and shaded area represent least squares linear fit with 95% CI. All reported r and p values are based on Pearson correlation.

**Acknowledgements**
We are grateful to Uri Hasson for guidance in the experimental design and analysis, and helpful comments on the manuscript

**Funding**
This study was funded by NCCIH Grant R01AT010627 and NIDA Grant R01DA048301

**Author contributions ([CRediT format](CRediT format))**
Greg Kronberg: Conceptualization, Methodology, Formal Analysis, Writing
Ahmet O. Ceceli: Conceptualization, Methodology, Formal Analysis, Writing
Natalie McClain: Investigation, Methodology
Devarshi Vasa: Investigation, Methodology
Yuefeng Huang: Investigation, Formal Analysis
Pierre-Olivier Gaudreault: Investigation, Formal Analysis, Visualization
Sarah King: Investigation, Formal Analysis
Pazia Miller: Project Administration, Investigation
Lily Gabay: Project Administration, Investigation
Pias Malaker: Project Administration, Investigation
Defne Ekin: Project Administration, Investigation
Nelly Alia-Klein: Conceptualization, Methodology, Writing, Supervision
Rita Z. Goldstein: Project Administration, Conceptualization, Methodology, Writing, Supervision

**Competing interests**
The authors declare no competing interests

**Data and materials availability**
Data and code will be made available upon request

**Supplementary Materials**
Table S1-S4
Fig. S1-S2
References 1-29
Movie S1-S4

**Supplemental Methods**

*Participants*

A total of 29 iOUD (40.4±10.3 years, 23 M) and 16 HC (43.8±10.3 years, 10 M) were recruited through advertisements, local treatment facilities, and word of mouth. All participants provided written informed consent. The study was approved by the Icahn School of Medicine at Mount Sinai's Institutional Review Board. Demographics and selected drug use variables of these participants are presented in Table S1. A comprehensive clinical diagnostic interview was conducted, comprised of the Mini International Neuropsychiatric Interview (MINI) (*22*) and the Addiction Severity Index (*23*) to assess the severity as well as recent and lifetime history of alcohol- and drug-related problems. The severity of drug dependence, craving, and withdrawal symptoms were determined using the Severity of Dependence Scale (*24*), Heroin Craving Questionnaire (a modified version of the Cocaine Craving Questionnaire) (*25*), and the Subjective Opiate Withdrawal Scale (*26*), respectively. In brief, all iOUD met criteria for heroin use disorder (primary route of administration: 14 intravenous, 11 nasal, 3 smoked/inhaled, 1 oral). Other comorbidities in iOUD included cocaine use disorder (n=9), antisocial personality disorder (n=4), alcohol use disorder (n=3), benzodiazepine use disorder (n=3), post-traumatic stress disorder (n=2), polysubstance use disorder (n=2), marijuana use disorder (n = 2), general anxiety disorder (n=2), persistent depressive disorder (n=2), major depressive disorder (n=1), methamphetamine use disorder (n=1), (another) opiate use disorder (n=1), and/or panic disorder (n=1). All of these other common comorbidities (*27, 28*) were either in partial or sustained remission at time of study. No comorbidities were found for the HC subjects. Exclusion criteria for all participants were the following: 1) DSM-5 diagnosis for schizophrenia or developmental disorder (e.g., autism); 2) Head trauma with loss of consciousness (>30 min); 3) History of neurological disease of central origin including seizures; 4) Cardiovascular disease including high blood pressure and/or other medical conditions, including metabolic, endocrinological, oncological or autoimmune diseases, and infectious diseases common in iOUD including Hepatitis B and C or HIV/AIDS; 5) Metal implants or other MR contraindications (including pregnancy). We did not exclude for DSM-5 diagnosis of a drug use disorder other than opiates as long as heroin was the primary drug of choice/reason for treatment-seeking, since iOUD commonly use alcohol, amphetamines, benzodiazepines and other sedatives, cocaine, and marijuana. Exclusion criteria for the HC were the same, except history of any drug use disorder was prohibitive. All iOUD were stabilized on medication (methadone or buprenorphine) as confirmed by urine assays (methadone = 24, buprenorphine = 5). All 29 iOUD, and one HC, were current cigarette smokers.

| | OUD (N = 29) | HC (N = 16) | Significance test |
|---|---|---|---|
| Age [a] | 40.4 ± 10.3<br>Min: 27.7, Max: 60.3 | 43.8 ± 10.3<br>Min: 30.7, Max: 58.5 | $t(43) = 1.13, p = .2661$ |
| Sex (M/F) [b] | 23/6 | 10/6 | $\chi2(1, N = 45) = .75, p = .385$ |
| Race (Black/White/Other) [b] | 2/19/8 | 5/11/0 | $\chi2(2, N = 45) = 5.72, p = .057$ |
| Ethnicity (Hispanic/non-Hispanic) [b] | 13/16 | 3/13 | $\chi2(1, N = 45) = 2.03, p = .154$ |
| Handedness (Left/Right) [b] | 3/26 | 4/12 | $\chi2(1, N = 45) = .75, p = .385$ |
| Education (years; 12=high school graduate) [c] | 12.3 ± 2.17<br>Min: 9, Max: 17 | 16.66 ± 3.04<br>Min: 13, Max: 25 | $W = 416.5, p < .001*$ |
| Verbal IQ (WRAT-III reading subscale) [a] | 93.3 ± 12.3<br>Min: 71, Max: 116 | 110.63 ± 8.29<br>Min: 93, Max: 123 | $t(43) = 5.05, p < .001*$ |
| Non-verbal IQ (WASI-II matrix reasoning) [c] | 10.6 ± 2.74<br>Min: 3, Max: 15, Missing: 1 | 12.31 ± 2.91<br>Min: 4, Max: 17 | $W = 313.5, p = .028$ |
| Beck Depression Inventory [c] | 12.9 ± 11.4<br>Min: 4, Max: 46 | 3.38 ±4.94<br>Min: 0, Max: 17 | $W = 89, p < .001*$ |
| Fagerstrom Test for Nicotine Dependence | 3.59 ± 1.52<br>Min: 0, Max: 7 | - | |
| Age at first heroin use | 24.7 ± 7.54<br>Min: 15, Max: 42 | - | |
| Age at regular heroin use | 25.5 ± 7.41<br>Min: 15, Max: 43 | - | |
| Lifetime heroin use (years) | 11.4 ± 7.89 | - | |

|  | Min: 2, Max: 27 |  |  |
|---|---|---|---|
| Period of heaviest use (years) | 5.07 ± 6.24<br>Min: 0.08, Max: 27, Unreported: 1 | - |  |
| Days since last heroin use | 196 ± 272<br>Min: 3, Max: 1007 | - |  |
| Days of heroin use in last 30 days | 0.31 ± 0.85<br>Min: 0, Max: 4 | - |  |
| Severity of Dependence Scale | 10.0 ± 3.58<br>Min: 0, Max: 15 | - |  |
| Heroin craving questionnaire | 43.6 ± 15.4<br>Min: 16, Max: 86, Unreported: 1 | - |  |
| Subjective Opiate Withdrawal Scale | 3.28 ± 3.40<br>Min: 0, Max: 12 | - |  |

*Table S1. Demographics and drug use variables*
Significant group differences after control for multiple comparison (p < 0.05/9) are flagged with an asterisk. Continuous variables are reported as mean ± standard deviation. [a] two-sample T test; [b] Chi-squared test; [c] Wilcoxon rank sum test. WRAT: Wide Range Achievement Test; WASI: Wechsler Abbreviated Scale of Intelligence, 2nd edition

*MRI acquisition and movie*
MRI scans were acquired with a Siemens 3T Skyra (Siemens, Erlangen, Germany) and a 32-channel head coil while participants passively viewed the first 17 min 3 sec of the movie "Trainspotting" using MRI-compatible in-ear headphones. The MRI protocol was optimized to be Human Connectome Project compatible (*29*, *30*). Blood oxygen level dependent (BOLD) fMRI responses were assessed via a T2*-weighted single-shot multi-band (acceleration factor of 7) gradient-echo EPI sequence (TE/TR=35/1000 ms), with 2.1 mm isotropic resolution, 70 axial slices for whole brain coverage (14.7 cm), 206 × 181 mm FOV, 96 × 84 matrix size, 60°-flip angle, blipped CAIPIRINHA phase-encoding shift=FOV/3, ~2 kHz/pixel bandwidth with ramp sampling, 0.68 ms echo spacing, and 57.1 ms echo train length. T1-weighted anatomical scans were acquired using a 3D MPRAGE sequence (TR/TE/TI=2400/2.07/1000 ms) with 0.8 mm isotropic resolution, 256 × 256 × 179 mm$^3$ FOV, 8° flip angle with binomial (1, −1) fat saturation, 240 Hz/pixel bandwidth, 7.6 ms echo spacing, and an in-plane acceleration (GRAPPA) factor of 2. The scan session included other procedures unrelated to the movie to be reported elsewhere. Immediately before and after the movie, subjects were asked to rate their desire for heroin on a scale of 0-9.

*Scene-specific post-movie survey*
Outside of the scanner (within 45 minutes after watching the movie), each subject completed scene-specific craving ratings. For this purpose, we extracted 3-sec clips separated by regular 30 sec intervals to densely sample movie scenes in an unbiased manner. This yielded 34 3-sec clips from 20 of the 24 total scenes in the 17-min movie (see *Scene designation and labeling*). Participants were instructed to watch each clip and provide subjective ratings of their experience when they watched that clip in the scanner, including their craving, and any scene-induced emotion amplification or suppression (for purposes of measuring savoring vs. reappraising, respectively, not related to the current goals). For correlations in Fig. 2, these scene-by-scene craving ratings were averaged over all 34 3-sec clips for each subject. The survey also included a question to determine whether the participant had previously watched the movie, self-report ratings of attention during movie-watching, understanding of the movie, ratings of perceived audio and video quality, followed by a four-question memory probe related to the general theme of the movie: "On which drug is the movie primarily centered?", "Is the main character trying to quit using a drug?", "Did the main character and his friends live in rich neighborhoods?", and "Did anyone in the movie die of drug-related causes, and if so, how many?". The survey also included a recognition task comprising a series of drug and non-drug word stimuli representing target and distractor items, where participants were asked whether they had seen each item in the movie, and how confident they felt about their response.

*Anatomical data preprocessing with fMRIprep version 20.2.1 (31)*
For each subject, a T1-weighted (T1w) image was corrected for intensity non-uniformity with N4BiasFieldCorrection (*32*), distributed with ANTs 2.3.3 (*33*), and used as T1w-reference throughout the workflow. The T1w-reference was then skull-stripped with a Nipype implementation of the antsBrainExtraction.sh workflow (from ANTs), using OASIS30ANTs as target template. Brain tissue segmentation of cerebrospinal fluid, white-matter and gray-matter was performed on the brain-extracted T1w using fast [FSL 5.0.9, RRID:SCR_002823 (*34*)]. Volume-based spatial normalization to a standard space (MNI152NLin2009cAsym) was performed through nonlinear registration with antsRegistration (ANTs 2.3.3), using brain-extracted versions of both T1w reference and the T1w template. The ICBM 152 Nonlinear Asymmetrical template version 2009c [(*35*), RRID:SCR_008796; TemplateFlow ID: MNI152NLin2009cAsym] was used for spatial normalization.

*Functional data preprocessing with fMRIprep version 20.2.1*

For all BOLD movie runs (across all subjects), the following preprocessing was performed. First, a reference volume and its skull-stripped version were generated by aligning and averaging a single-band reference. A B0-nonuniformity map (or fieldmap) was estimated based on two echo-planar imaging (EPI) references with opposing phase-encoding directions, with AFNI's 3dQwarp (*36*). Based on the estimated susceptibility distortion, a corrected EPI (echo-planar imaging) reference was calculated for a more accurate co-registration with the anatomical reference. The BOLD reference was then co-registered to the T1w reference using flirt [FSL 5.0.9, (*37*)] with the boundary-based registration (*38*) cost-function. Co-registration was configured with nine degrees of freedom to account for distortions remaining in the BOLD reference. Head-motion parameters with respect to the BOLD reference (transformation matrices, and six corresponding rotation and translation parameters) were estimated before any spatiotemporal filtering using mcflirt [FSL 5.0.9, (*39*)]. First, a reference volume and its skull-stripped version were generated using a custom methodology of fMRIprep. The BOLD time-series were resampled onto their original, native space by applying a single, composite transform to correct for head-motion and susceptibility distortions. The BOLD time-series were resampled into the MNI152NLin2009cAsym standard space and used for further custom preprocessing.

*Custom preprocessing*

Spatially-normalized, preprocessed BOLD data in the MNI152NLin2009cAsym standard space were further processed with the following steps. A gray-matter mask was generated by averaging tissue probability maps of all subjects, then binarizing the resulting map by thresholding at 95% probability gray-matter. The gray-matter mask and gaussian smoothing (6 mm) were then applied. In the remaining gray-matter voxels, the first 10 TR's were removed from all BOLD time series to eliminate contributions of large stimulus onset responses to further analyses (*5*). The following confounds were regressed out of each BOLD signal: 6 translation and rotation parameters (x,y,z for each), their square, their derivative, and their squared derivative, as well as the global cerebrospinal fluid component output by fmriprep. Regression of confounds, high-pass filtering (period of 140 s), linear detrending, and z-scoring were applied in a single step using the signal.clean function from nilearn python package (*40*).

*Global component and selective components*

For each subject the preprocessed BOLD time series were averaged over all gray-matter voxels and then z-scored, resulting in a single global component per subject. Selective components were then derived by regressing the time series at each voxel within a subject onto the global component for that subject with a least-squares linear model. The residual of this linear fit was then z-scored and kept as the selective component for that subject-voxel (*6*). For region of interest based analyses, the selective components at all voxels within a mask were first averaged within a subject and then z-scored again to be used as the region-specific component in further analyses.

*Identifying synchronized TR's and region-specific reactivity*

For a given group and region of interest (or global component), and employing a single time series (1019 TR's) per subject, a sliding one-sample t-test with a threshold of $p<0.01$ was used to find the TR's where the group mean was significantly different from zero (*6*). Significant tests yielded a set of region and group-specific time points, $T^*=\{t_i\}$, interpreted to reflect the most synchronized activity across the group of subjects in a specific region of interest. Once $T^*$ was determined for a group, region-specific reactivity for each subject was calculated by taking the mean of their region-specific component sampled at $T^*$ only. This subject and region-specific reactivity score was then used as a regressor for further correlation analyses with behavioral variables.

*Reverse correlation and region-specific movies*

To gain insight into the movie content that drove the synchronization in a group of subjects, we performed a reverse correlation-like analysis (inspired by analysis of single unit recordings) as in Hasson et al. (*6*). For each TR in $T^*$, we collected a 5 second clip of the movie corresponding to $t_i-10$ to $t_i-5$ to account for the delay in the hemodynamic response, which typically peaks in 4-6 sec and takes 10-12 sec to decay to baseline in the NAc during event-related fMRI tasks (*41*). Individual clips for each $t_i$ were then sorted by the p-value from the t-test associated with $t_i$ and concatenated into a single "region-specific movie", which was used for further analysis.

*Labeling of region-specific movies*

The movie clip was separated into 24 scenes based on shifts in narrative [e.g., location, theme, and time; consistent with Chen et al. (*11*)] by two independent raters who were naïve to the analyses. Raters then categorized each scene as "drug" or "non-drug" depending on whether they contained visual drug cues such as the presence of drugs or drug use, yielding 9 drug and 15 non-drug scenes (506 and 517 total seconds, respectively). To analyze the content of region-specific movies, we collected all TR's contained within a region-specific movie and matched each TR to one of the 24 identified scenes. Each TR was then assigned a label as either drug or non-drug based on the label of the scene that it belonged to. This yielded a distribution of scene-type labels for each region-specific movie. We first tested for a main effect of scene-type against the null hypothesis that $T^*$ was selected at random. This would lead to distributions of scene type that are proportional to the total time (in seconds) that they occupied in the movie, i.e., under the null hypothesis of random $T^*$, we expected to find 49.46% and 50.54% of drug and non-drug scenes, respectively (i.e., 506 and 517 out of 1023 seconds total). We combined the scene-type counts for both groups and used a one sample chi-square to test if this distribution of scenes was significantly different from the expected distribution. We similarly tested for a main effect of the group, with a one-sample chi-square test and a null hypothesis of equal counts for both groups. Finally, we tested for an interaction between group and scene-type using a chi-square independence test on the 2x2 (scene-type x group) contingency table.

**Supplemental Results**

We tested whether the LNAc reactivity was related to other potential explanatory variables including the demographic measures that differed between the groups and, in iOUD, measures of recent and lifetime drug use, withdrawal symptoms and severity of dependence (Table S2)

|  | **OUD LNAc-reactivity r, *p*** | **HC LNAc-reactivity r, *p*** | **OUD+HC combined LNAc-reactivity r, *p*** |
|---|---|---|---|
| Education (years; 12=high school graduate) | 0.31, 0.11 | 0.13, 0.64 | -0.10, 0.50 |
| Verbal IQ (WRAT-III) | 0.31, 0.11 | 0.30, 0.27 | -0.12, 0.42 |
| Beck Depression Inventory | -0.16, 0.43 | 0.44, 0.09 | 0.26, 0.09 |
| Fagerstrom Test for Nicotine Dependence | 0.23, 0.24 | | |
| Age at first heroin use | 0.11, 0.58 | | |
| Age at regular heroin use | 0.05, 0.82 | | |
| Lifetime heroin use (years) | 0.05, 0.80 | | |
| Period of heaviest use (years) | 0.31, 0.12 | | |
| Days since last heroin use | -0.20, 0.31 | | |
| Days of heroin use in last 30 days | 0.06, 0.78 | | |
| SDS | -0.09, 0.64 | | |
| HCQ | 0.23, 0.25 | | |
| SOWS | 0.26, 0.18 | | |

*Table S2. No significant Pearson correlation between LNAc-reactivity and other potential explanatory variables*
WRAT: Wide Range Achievement Test; SDS: Severity of Dependence Scale; HCQ: Heroin Craving Questionnaire; SOWS: Subjective Opiate Withdrawal Scale.

*Nonparametric test of scene types in LNAc-specific movies*

To further test the statistical significance of the drug-biased reactivity observed in the LNAc in iOUD, we generated a null distribution for the number of drug-scenes in the LNAc-specific movie. For each subject, the mean LNAc signal was phase-randomized 5000 times and reverse correlation analyses were repeated to get a count of drug scenes for each randomization. This yielded a null distribution for the number of drug scenes in the LNAc-specific movie for a completely random signal. The LNAc-specific movie had a significant number ($\alpha$=0.05) of drug scenes in iOUD, but not in HC.

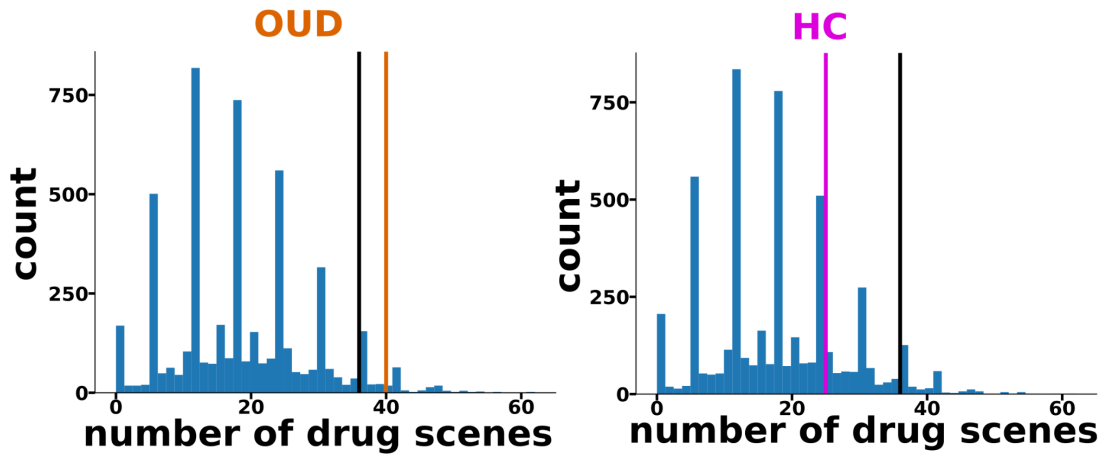

*Fig. S1. Comparison of LNAc-specific movies to phase-randomized signals*
Null distributions for the number of drug scenes in a LNAc-specific movie are plotted in blue (5000 independent phase-randomizations). Black vertical lines indicate the 95th percentile of the 5000 phase-randomizations. The actual number of drug scenes for OUD (left, orange vertical line) is above this threshold, but not for HC (right, magenta vertical line).

*Comparison of LNAc-specific components between groups*

To better understand the relationship between the LNAc signal in each group, we directly compared the LNAc-specific component between groups at each set of significant time points (significant peaks or valleys derived from each group). Taking the mean over all significant time points, HC and iOUD were significantly different at each set of time points ($p<0.05$, two-sample t-test), highlighting the differences for the LNAc signal between the groups.

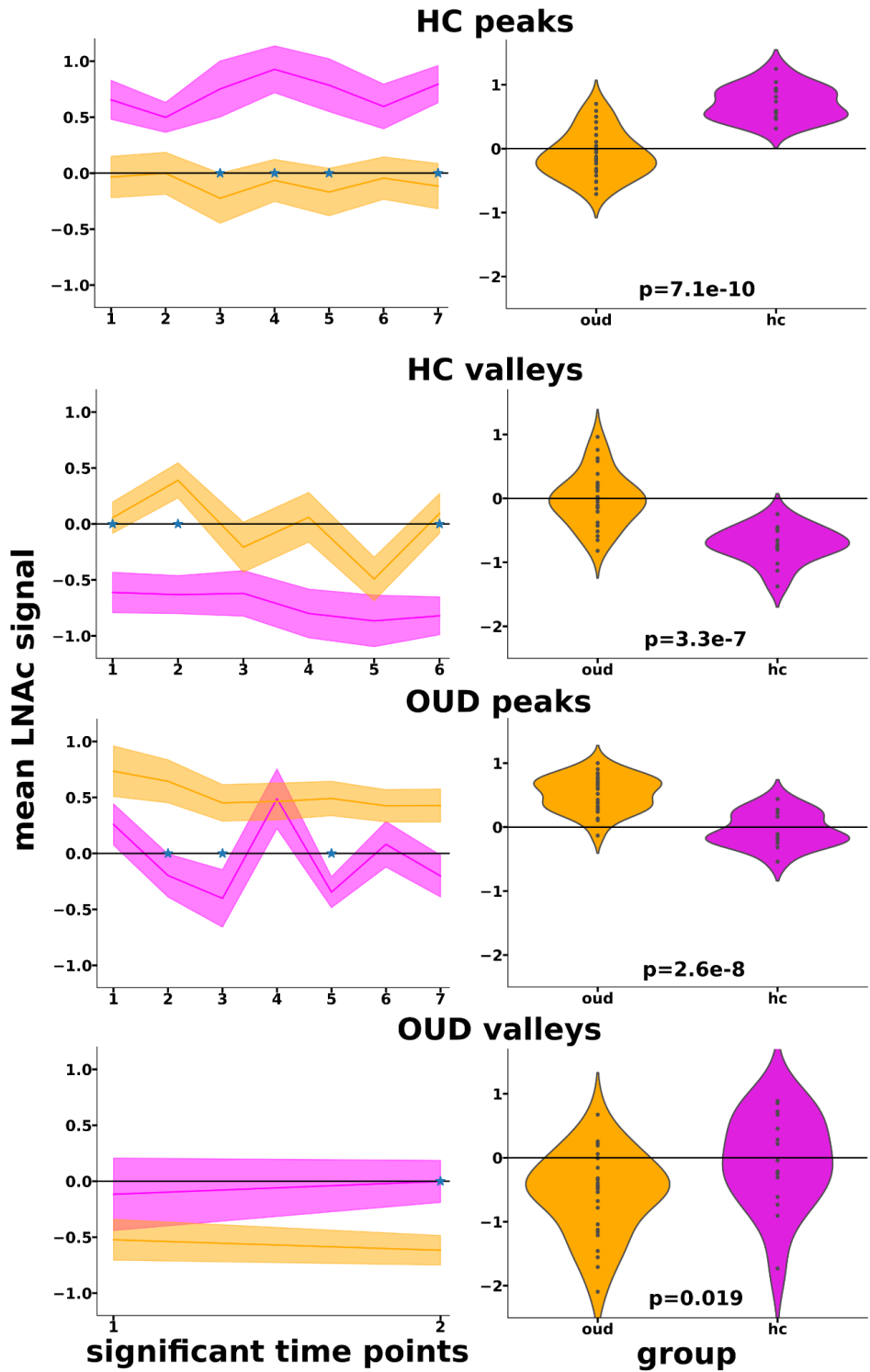

*Fig. S2. Comparison of LNAc signal between groups*
For each group, significant time points in the LNAc signal ($p<0.05$ one-sample t-test) were split into peaks and valleys, yielding 4 sets of time points (peaks/valleys x OUD/HC). The mean signal at each set of time points is compared between groups (left) as well as the mean over all time points (right) (iOUD=orange/HC=magenta). Time points that are significantly different between the groups ($p<0.05$, two-sample t-test) are labeled with blue stars (left).

*Reverse correlation of global component/other regions*

      To assess the specificity of our results to the LNAc, we repeated our analyses for the ventromedial prefrontal cortex, dorsolateral prefrontal cortex, anterior cingulate cortex, anterior insula, and putamen, regions that have previously been associated with drug cue-reactivity (*3*) and the fusiform cortex as a control region. We extracted signal from these regions using independent coordinates from previous studies (*13*, *15*, *42*, *43*) as well as anatomical atlases. Table S3 lists all regions that were tested, along with the results of a chi-square contingency test for the group by scene-type interaction effect. Only the LNAc and LaInsula region-specific movies showed a significant interaction between group and scene-types (p<0.05, Bonferroni correction with 15 tests). However, only the LNAc showed a complete reversal of scene-type preferences between groups (i.e., drug scenes for OUD and non-drug scenes for HC).

| | | | | OUD | | HC | | | |
|---|---|---|---|---|---|---|---|---|---|
| **region** | x | y | z | Drug | Non-drug | Drug | Non-drug | $\chi^2$ (1) | *p* |
| L rvACC (*44*) | -9 | 54 | -11 | 30 | 17 | 41 | 12 | 1.61 | 0.21 |
| R rvACC [a] | 9 | 54 | -11 | 26 | 13 | 42 | 12 | 0.91 | 0.34 |
| L vmPFC (*13*) | -3 | 42 | -12 | 30 | 20 | 42 | 19 | 0.60 | 0.44 |
| R vmPFC [a] | 3 | 42 | -12 | 42 | 14 | 48 | 20 | 0.12 | 0.73 |
| L dlPFC (*42*) | -30 | 36 | 42 | 38 | 16 | 12 | 12 | 2.18 | 0.14 |
| R dlPFC [a] | 30 | 36 | 42 | 36 | 30 | 14 | 23 | 2.02 | 0.15 |
| L aInsula (*43*) | -41 | 21 | 3 | 67 | 14 | 30 | 22 | 8.82 | 0.0030* |
| R aInsula [a] | 41 | 21 | 3 | 62 | 28 | 42 | 26 | 0.59 | 0.44 |
| L NAc [b] | -10 | 12 | -8 | 40 | 8 | 25 | 40 | 20.95 | <0.00001* |
| R NAc [b] | 9 | 13 | -7 | 34 | 18 | 44 | 24 | 0.013 | 0.91 |
| L Putamen [b] | -25 | 1 | 1 | 18 | 48 | 18 | 30 | 0.91 | 0.34 |
| R Putamen [b] | 25 | 2 | 0 | 42 | 36 | 18 | 19 | 0.10 | 0.74 |
| L Fusiform [b] | 36 | -25 | -28 | 14 | 39 | 26 | 48 | 0.72 | 0.40 |
| R Fusiform [b] | -33 | -53 | -17 | 123 | 60 | 128 | 71 | 0.54 | 0.46 |
| Global component | -- | -- | -- | 24 | 18 | 33 | 10 | 4.91 | 0.027 |

*Table S3. Scene-types in region-specific movies for other drug cue-reactivity regions and control regions*
Chi-square statistics and p-values are for contingency tests between group and scene-type. The table is sorted along the y axis, anterior to posterior. rvACC: rostroventral anterior cingulate cortex; vmPFC: ventromedial prefrontal cortex; dlPFC: dorsolateral prefrontal cortex; aInsula: anterior insula; NAc: nucleus accumbens.
[a]Region of interest masks derived by x-inverted coordinates from the corresponding region's contralateral hemisphere
[b]anatomically-derived region of interest masks using the Harvard-Oxford Cortical/Subcortical Atlas (50% threshold)

*Probing movie comprehension*
The OUD and HC groups did not significantly differ in any of the post-movie survey measures and we did not exclude any subjects on this basis.

|  | OUD | HC | *Significance test* |
|---|---|---|---|
| Previously seen movie? (Yes/No): | 10/19 | 8/8 | $\chi2(1, N = 45) = .489, p = .484$ |
| Self-reported attention during movie (0-9) | 6.24 (2.01) | 6.94 (2.38) | $W = 288, p = .181$ |
| Self-reported understanding of the movie (0-9) | 5.97 (2.68) | 6.00 (2.61) | $W = 230, p = .981$ |
| Subjective movie audio quality rating (0-9) | 3.10 (2.55) | 2.50 (2.53) | $W = 196, p = .401$ |
| Subjective movie video quality rating (0-9) | 6.69 (2.61) | 5.69 (2.70) | $W = 173, p = .161$ |
| Proportion of memory probe questions answered correctly | 0.73 (0.19) | 0.77 (0.23) | $W = 259, p = .495$ |
| Proportion of drug items correctly recognized | 0.74 (0.12) | 0.76 (0.08) | $W = 241, p = .844$ |
| Confidence about drug item recognition (0-9) | 7.05 (2.12) | 5.98 (2.05) | $W = 151, p = .056$ |
| Proportion of non-drug items correctly recognized | 0.81 (0.14) | 0.83 (0.07) | $W = 235, p = .951$ |
| Confidence about non-drug item recognition (0-9) | 6.30 (2.64) | 5.98 (2.27) | $W = 207, p = .561$ |

*Table S4. Movie comprehension and memory did not significantly differ between groups*
Data are reported as group mean (SD), unless otherwise indicated. Significance tests reflect Wilcoxon rank sum tests, except for "previously seen movie?" (chi-square test).

**Supplemental videos**
video S1 - OUD peaks in LNAc
video S2 - OUD valleys in LNAc
video S3 - HC peaks in LNAc
video S4 - HC valleys in LNAc